\documentclass[amsmath,amssymb,aps]{revtex4} 
\usepackage[dvipdfmx]{graphicx}
\usepackage{amsmath}
\usepackage{bm}

\begin{document}

\title{Nambu-Goldstone Gravity and New Cosmology}

\author{Kimihide Nishimura}
\email{kimihiden@dune.ocn.ne.jp}
\affiliation{Nihon-Uniform, 1-4-1 Juso-Motoimazato, Yodogawa-ku Osaka 532-0028 Japan}
\date{\today}

\begin{abstract}
The reformulation of gravity as the Nambu-Goldstone mode associated with broken gauge symmetry is given and applied to cosmology. We find that; the new gravity consists of a scalar mode and a tensor mode, of which the tensor mode corresponds to the Einstein gravity; the universe pulsates without spacial curvature; the gravitational constant increases from zero as the universe expands and decreases to zero as the universe contracts;  
the dark energy is identified with the sum of the gravitational energies of scalar and tensor modes, 
while the dark matter is the mirage of the gravitational energy caused by the projection into the $\Lambda$CDM spectrum;  
the gravitational equation dictates the cosmological constant to be zero;  
the universe begins with inflation, and the present age of the universe is approximately 550Gy.
\end{abstract}

\maketitle

\section{Introduction}
The recent discoveries in cosmology by the advance of observational technologies; 
1. the necessity of dark energy, dark matter, and inflation for understanding in particular 
the accelerating expansion of the universe\cite{Riess0,Perlmutter} and 
the anisotropies of cosmic microwave background\cite{PlanckCollaboration,Hinshaw,Riess}; 
2. the abundance of matured galaxies with high redshifts found by the James Webb Space Telescope (JWST), 
which possibly contradicts the present age of the universe\cite{IvoLabbe,Carnall}, suggest that the Einstein theory of gravity has already been inadequate for describing our universe. In order to accommodate these new entities, it seems necessary to revise the theory of gravity. 

From theoretical point of view, one of the unsatisfactory points with Einstein gravity is that it is not based on quantum theory. The equivalence principle, on which Einstein gravity is based, alone is not adequate to determine the dynamics of gravity, though the Einstein-Hilbert action is the simplest one. 

Another distinction of gravity from other gauge interactions is the weakness of the force. Dirac \cite{Dirac} once tried to understand the small gravitational constant by supposing that it decreases as the universe expands. 
However, a quantum theoretical reconsideration on this problem suggests the opposite possibility: the gravitational constant increasing from zero as the universe expands. This possibility is realized when gravity is considered as the Nambu-Goldstone (NG) mode associated with broken symmetry \cite{NJ,Goldstone,GSW,KN}. If the universe begins with spontaneous symmetry breakdown, there would be no NG bosons before the breakdown, and therefore no gravity. Further, if the breakdown is not yet complete, gravity would be weak. 
Then, the hypothesis of NG gravity transfers the problem to finding out the symmetry which breaks down at the beginning of the universe, which will in turn provide a missing principle to determine the dynamics of gravity. 
Here we emphasize that what is required to be broken is not spacetime symmetry.

The hypothetical necessity of spacetime symmetry breakdown seems to stem from the desire to understand spinning massless particles as the NG bosons, for example, photons and gravitons \cite{Bjorken}. Of those, the idea of NG photon was not so attractive, since there already exists gauge theory. On the other hand, the NG graviton is far more important, since it enable us to understand gravity, which is otherwise based on the utterly different principle, in the context of gauge theory.

Spontaneous symmetry breakdown, which plays the essential role for the unified picture of elementary particles and forces\cite{Higgs,Weinberg0}, postulates that some quantum field develops the vacuum expectation value, 
which is supposed to be the same at all spacetime points. 
Considering this phenomenon in the presence of gravity, the vacuum expectation value independent of spacetime points seems too restrictive. 

However, according to the equivalence principle,  
we can find at every spacetime point a local inertial frame where gravity is absent, and where spacetime symmetry holds locally. This observation suggests that, even in the presence of gravity, we may still assume a common vacuum expectation value at every local Lorentz frame. In addition, it makes aware us an important fact that the local Lorentz frame transforms spacetime symmetry (global Lorentz symmetry) into gauge symmetry (local Lorentz symmetry), which is not accompanied by any spacetime coordinate transformations. 

Hitherto in elementary particle physics the fields which break gauge invariance are supposed to be Lorentz scalars in order not to break Lorentz invariance.  The reasoning above will remove this restriction, if we consider symmetry breaking in a local Lorentz frame. Then, any local Lorentz tensors with gauge indices become allowable as the symmetry breaking fields, which do not break the general coordinate invariance nor the equivalence principle. 

When the gauge-breaking field is a local Lorentz tensor, its vacuum expectation value will take the form of a constant mixed tensor with both local Lorentz indices and the gauge indices. 
If we want to preserve local Lorentz invariance even after symmetry breakdown, the constant mixed tensor should have the form of some multiple of the Minkowski metric. 
This requirement implies that the gauge group to be broken should have as a subgroup Lorentz group of four dimensions, or it should be some extension of Lorentz group, for example, a Lorentz group of higher dimensions. 
We soon recognize that five dimensional Lorentz symmetry is adequate for generating gravity as NG mode. 
In this case, no extra symmetry other than the ordinary Lorentz symmetry survives. 
Actually, as presently shown, the vierbein, which is introduced as an auxiliary field to set up a local Lorentz frame, 
acquires dynamics by spontaneous symmetry breakdown in a local Lorentz frame.  

The simplest choice of the gauge-breaking tensor field is to take the gauge field itself. 
We construct in Sec.\ref{LNG} the action of NG gravity explicitly based on the Yang-Mills gauge field with five dimensional local Lorentz symmetry on four dimensional Minkowski spacetime.
The resulting action contains a scalar field and the quadratic term of the Riemann curvature, which are both absent in the Einstein theory, where one of the interaction terms of the scalar field becomes the germ of the Einstein action of gravity.

Section \ref{GEQ} derives the gravitational equations for the scalar and the tensor modes. 
The gravitational constant is determined by the vacuum expectation value of the scalar field, 
as once supposed by Brans and Dicke \cite{BransDicke}. 
The new gravitational equation of tensor mode is expressible as the total energy-momentum tensor to be zero, 
which implies that there remains no role for the cosmological constant except calibrating the zero points of the constituent energy densities.

Sec.\ref{FEQ} applies the new gravity to the expanding universe to obtain the Friedmann equations. 

Sec.\ref{ESS} shows that the Friedmann equations thus derived admit one parameter family of exact solutions, 
where the scalar gravity determines the time evolution of the universe, while the tensor gravity determines the energy spectrum in the universe. 
The solutions show that the universe pulsates without spacial curvature, and begins with the power law inflation.  

Sec.\ref{DEDM} shows that the dark energy and the cold dark matter in the Big-Bang cosmology are both originated from the same gravitational energy obeying different equations of state, though the cold dark matter is simply the reflection of the gravitational energy converted into the state of cold matter.

Sec.\ref{VES} clarifies that the cause of the phase transition of the universe from inflationary expansion to decelerating expansion is the change of the nature of gravity from repulsive into attractive. 

Sec.\ref{CMB} calculates the temperatures of thermal radiation and normal matter to reconsider 
the cosmic microwave background (CMB), the Big-Bang nucleosynthesis (BBN), and the baryon asymmetry (baryogenesis) in the context of new cosmology. 

Sec. \ref{VGC} determines the free parameter remaining until the end in our cosmological model, 
and estimates the present age of the universe, by comparing the rate of change of the gravitational constant with observational data.

Sec.\ref{GRP} represents the results obtained in graphics.
\section{New Lagrangian of gravity \label{LNG}}
We first construct the Yang-Mills action with five dimensional local Lorentz gauge invariance in four dimensional Minkowski spacetime. 
The simplest presentation is to consider the scalar multiplet $V_A$ with gauge index $A=(\alpha,5)=(0,1,2,3,5)$, which transforms as a local Lorentz vector in the gauge space,  
\begin{equation}
\delta V_A=\theta_A{}^BV_B,
\end{equation}
where $\theta_{AB}=-\theta_{BA}$ are the gauge parameters for the extended local Lorentz transformations. 
A subscript or a superscript is interchangeable by contracting it with the five dimensional Minkowski metric 
$\eta_{AB}=\eta_{BA}$ of signature $(+,-,-,-,-)$.  
The gauge connection $\omega_{\mu AB}=-\omega_{\mu BA}$ is introduced to define the covariant derivative,
\begin{equation}
D_\mu V_A=\partial_\mu V_A+\omega_{\mu A}{}^BV_B,
\end{equation}
where $\mu$ is the coordinate index of four dimensional Minkowski spacetime: $\mu=(0,1,2,3)$.
The gauge covariance requires that $\omega_{\mu AB}$ transforms as
\begin{equation}
\delta\omega_{\mu AB}=\theta_A{}^C\omega_{\mu CB}+\theta_B{}^C\omega_{\mu AC}-\partial_\mu\theta_{AB}.
\end{equation}
The curvature tensor, or the field strength $\Omega_{\mu\nu AB}$ is obtained from the relation:
\begin{equation}
[D_\mu, D_\nu]V_A=\Omega_{\mu\nu A}{}^BV_B,\quad 
\Omega_{\mu\nu AB}=\partial_\mu\omega_{\nu AB}-\partial_\nu\omega_{\mu AB}
+\omega_{\mu A}{}^C\omega_{\nu CB}-\omega_{\nu A}{}^C\omega_{\mu CB}.
\end{equation}
Then, the Yang-Mills Lagrangian in the general coordinate system has the form 
\begin{equation}
{\cal L}_\omega=-\frac{1}{8g^2}g^{\mu\rho}g^{\nu\sigma}\Omega_{\mu\nu}{}^{AB}\Omega_{\rho\sigma AB},
\end{equation}
where $\mu$, $\nu$, etc. are, from now on, the general coordinate indices, 
while the original Minkowski coordinate index $\mu$ is transferred to the local Lorentz index $\alpha$ by introducing the vierbein $e_\mu{}^\alpha$ as an auxiliary field. 
The metric tensor $g_{\mu\nu}$ is then expressed as $g_{\mu\nu}=\eta^{\alpha\beta}e_{\mu\alpha}e_{\nu\beta}$, and $g$ is a dimensionless gauge coupling constant. 

We next decompose the five dimensional local Lorentz tensor into 
the four dimensional local Lorentz components and the others by introducing notations: 
$\omega_{\mu\alpha}:=\omega_{\mu\alpha5}=-\omega_{\mu\alpha}{}^5$, and 
$\Omega_{\mu\nu\alpha}:=\Omega_{\mu\nu\alpha5}=-\Omega_{\mu\nu\alpha}{}^5$ 
to obtain
\begin{eqnarray}
{\cal L}_\omega&=&\frac{1}{4g^2}g^{\mu\rho}g^{\nu\sigma}\Omega_{\mu\nu}{}^\alpha\Omega_{\rho\sigma\alpha}
-\frac{1}{8g^2}g^{\mu\rho}g^{\nu\sigma}\Omega_{\mu\nu}{}^{\alpha\beta}\Omega_{\rho\sigma\alpha\beta},\label{L_Omega}\\
\Omega_{\mu\nu\alpha}&=&\nabla_\mu\omega_{\nu\alpha}-\nabla_\nu\omega_{\mu\alpha},
\quad \nabla_\mu\omega_{\nu\alpha}:=\partial_\mu\omega_{\nu\alpha}
-\Gamma^\rho{}_{\mu\nu}\omega_{\rho\alpha}+\omega_{\mu\alpha}{}^\beta\omega_{\nu\beta},
\label{LLCD}\\
\Omega_{\mu\nu\alpha\beta}&=&\partial_\mu\omega_{\nu\alpha\beta}-\partial_\nu\omega_{\mu\alpha\beta}
+\omega_{\mu\alpha}{}^\gamma\omega_{\nu\gamma\beta}-\omega_{\nu\alpha}{}^\gamma\omega_{\mu\gamma\beta}
+\omega_{\mu\alpha}\omega_{\nu\beta}-\omega_{\nu\alpha}\omega_{\mu\beta}.
\end{eqnarray}
Since the Lagrangian has local Lorentz symmetry of four dimensions as the subgroup, 
the component $\omega_{\mu\alpha\beta}$ in $\omega_{\mu AB}$ is identical with the ordinary local Lorentz connection, 
and therefore, the derivative $\nabla_\mu$ in (\ref{LLCD}) is covariant under both the general coordinate transformations and the four dimensional local Lorentz transformations. 
The Christoffel symbol $\Gamma^\rho{}_{\mu\nu}$ is understood to be calculated for the metric $g_{\mu\nu}$, though it is cancelled in $\Omega_{\mu\nu\alpha}$.

We then break the five dimensional local Lorentz symmetry down to four dimensions in a local Lorentz frame 
by assuming that $\omega_{\mu\alpha}$ develops the vacuum expectation value of the form: 
\begin{equation}
\langle e^\mu{}_\alpha \omega_{\mu\beta}\rangle=\frac{g}{\sqrt{3}}\langle \phi\rangle\eta_{\alpha\beta},
\label{SLLSV}
\end{equation}
where $\phi$ is a real scalar field of mass dimension one, and $\eta_{\alpha\beta}$ is the ordinary four dimensional Minkowski metric in the local Lorentz frame. The normalization factor has been chosen for later convenience. 
The vacuum expectation value (\ref{SLLSV}) preserves the ordinary local Lorentz invariance 
and the general coordinate invariance.
What are broken are the extra four generators of the five dimensional local Lorentz group.

From the above observations, the following parametrization of $\omega_{\mu\alpha}$ is possible \cite{KN},
\begin{equation}
\omega_{\mu\alpha}=\frac{g}{\sqrt{3}}\phi e_{\mu\alpha},
\label{parametrization}
\end{equation}
since the vierbein has been introduced as an auxiliary field. 
This form satisfies automatically the condition for the vacuum expectation value (\ref{SLLSV}).
The desired symmetry breakdown is realized by introducing to the Lagrangian (\ref{L_Omega}) the breaking term:
\begin{equation}
{\cal L}_{\rm sb}=\frac{\eta^2}{2}g^{\mu\nu}\omega_\mu{}^\alpha\omega_{\nu\alpha}-\Lambda_0, \quad \Lambda_0:=\frac{g^2\eta^4}{3},
\label{SBT}
\end{equation}
where $\eta$ is a constant with mass dimension one, which gives the scale of symmetry breaking. 
The constant term $\Lambda_0$, which is unnecessary for symmetry breaking, has been introduced tentatively as the standard cosmological constant. 
The implication of the cosmological constant in the new gravity formalism, which is one of the main themes of this paper, will be discussed later.

If the torsion constraint: $\nabla_\mu e_{\nu\alpha}-\nabla_\nu e_{\mu\alpha}=0$ is imposed on the vierbein, 
the total Lagrangian ${\cal L}_g={\cal L}_\omega+{\cal L}_{\rm sb}$ is reduced to the form:
\begin{equation}
{\cal L}_g=\frac{1}{2}g^{\mu\nu}\partial_\mu\phi\partial_\nu\phi-V(\phi)-\frac{\phi^2}{6}R
-\frac{1}{8g^2}R^{\mu\nu\rho\sigma}R_{\mu\nu\rho\sigma}, \quad
V(\phi)=\frac{g^2}{3}(\phi^2-\eta^2)^2,
\label{LagOfGravity}
\end{equation}
where $R_{\mu\nu\rho\sigma}$ is the Riemann curvature, and $R=e^{\mu\alpha}e^{\nu\beta}R_{\mu\nu\alpha\beta}$ is the scalar curvature. 
Owing to the standard cosmological constant $\Lambda_0$, 
the minimum of the potential $V(\phi)$ becomes zero at $\phi=\pm\eta$. 
We adopt Lagrangian ${\cal L}_g$ for the action of ``Nambu-Goldstone gravity".
\section{Gravitational equations\label{GEQ}}
The gravitational equation in the presence of matter is derived by adding to ${\cal L}_g$ 
the normal matter Lagrangian ${\cal L}_n$ corresponding to leptons, quarks, gauge bosons, the Higgs boson, 
and the others which may include yet unknown particles and sources of energies besides gravity, 
and varying the total action with respect to $\phi$ and $g_{\mu\nu}$. 
Then, we obtain the new gravitational equations: 
\begin{eqnarray}
\left(\nabla^\rho\nabla_\rho+\frac{R}{3}\right)\phi+\frac{\partial V}{\partial\phi}&=&j_n,\label{SGEQ}\\
R^{\mu\alpha\beta\gamma}R^\nu{}_{\alpha\beta\gamma}
-\frac{g^{\mu\nu}}{4}R^{\alpha\beta\gamma\delta}R_{\alpha\beta\gamma\delta}
+2\nabla_\rho\nabla_\sigma R^{\mu\rho\nu\sigma}&=&2g^2(T^{\mu\nu}_\phi+T^{\mu\nu}_n),\label{TGEQ}
\end{eqnarray} 
where
\begin{eqnarray}
j_n&:=&\frac{\delta{\cal L}_n}{\delta\phi}, \\ 
T^{\mu\nu}_n&:=&-\frac{2}{\sqrt{-\vert g\vert}}\frac{\delta(\sqrt{-\vert g\vert}{\cal L}_n)}{\delta g_{\mu\nu}},\\
T^{\mu\nu}_\phi&:=&\nabla^\mu\phi\nabla^\nu\phi
-g^{\mu\nu}\left(\frac{1}{2}\nabla^\rho\phi\nabla_\rho\phi-V\right)
+\frac{1}{3}\left(\nabla^\mu\nabla^\nu\phi^2-g^{\mu\nu}\nabla^\rho\nabla_\rho\phi^2\right)
-\frac{\phi^2}{3}\left(R^{\mu\nu}-\frac{g^{\mu\nu}}{2}R\right).
\label{T_phi}
\end{eqnarray} 
The new gravitational equation for the tensor mode (\ref{TGEQ}) is the fourth oder partial differential equation 
with the dimensionless gauge coupling constant $g$. 
We can verify that 
\begin{eqnarray}
\nabla_\mu T^{\mu\nu}_\phi&=&j_n\nabla^\nu\phi, \label{DT_phi}\\
\nabla_\mu T^{\mu\nu}_n&=&-j_n\nabla^\nu\phi, \label{DT_n}\\
\nabla_\mu T^{\mu\nu}_t&=&0, \quad 
T^{\mu\nu}_t:=\frac{1}{2g^2}\left[-R^{\mu\alpha\beta\gamma}R^\nu{}_{\alpha\beta\gamma}
+\frac{g^{\mu\nu}}{4}R^{\alpha\beta\gamma\delta}R_{\alpha\beta\gamma\delta}
-2\nabla_\rho\nabla_\sigma R^{\mu\rho\nu\sigma}\right].
\label{T_t}
\end{eqnarray}
The equation (\ref{DT_phi}) follows from (\ref{SGEQ}) and (\ref{T_phi}), while (\ref{DT_n}) follows from (\ref{DT_phi}) and 
the gravitational equation (\ref{TGEQ}), since it requires $\nabla_\mu(T^{\mu\nu}_\phi+T^{\mu\nu}_n)=0$.  
The conservation of the energy-momentum tensor for tensor gravity (\ref{T_t}) follows from the Bianchi identity alone. If the $\phi$-dependence of ${\cal L}_n$ is negligible ($j_n=0$), 
then $\nabla_\mu T^{\mu\nu}_\phi=\nabla_\mu T^{\mu\nu}_n=0$.

Owing to the notation $T^{\mu\nu}_t$, the tensor equation is expressible as the total energy-momentum tensor to be zero,
\begin{equation}
T^{\mu\nu}=T^{\mu\nu}_\phi+T^{\mu\nu}_t+T^{\mu\nu}_n=0.
\label{ZEMT}
\end{equation}
The Einstein equation is obtained by decomposing $T^{\mu\nu}_\phi$ into $T^{\mu\nu}_s$ and $T^{\mu\nu}_c$, where
\begin{equation}
T^{\mu\nu}_\phi=T^{\mu\nu}_s-T^{\mu\nu}_c, \quad T^{\mu\nu}_c:=\phi^2\left(R^{\mu\nu}-\frac{g^{\mu\nu}}{2}R\right).
\label{CriticalT}
\end{equation}
Then, we have 
\begin{equation}
R^{\mu\nu}-\frac{g^{\mu\nu}}{2}R=8\pi G(T^{\mu\nu}_g+T^{\mu\nu}_n), 
\quad T^{\mu\nu}_g:=T^{\mu\nu}_s+T^{\mu\nu}_t, 
\label{EEG}
\end{equation}
where
\begin{equation}
G=\frac{1}{8\pi\phi^2},
\label{NewtonConstant}
\end{equation}
is the Newton ``constant".
If $j_n=0$, we have
\begin{equation}
\nabla_\mu T^{\mu\nu}_n=0, \quad 
\nabla_\mu T^{\mu\nu}_g=\left(R^{\mu\nu}-\frac{g^{\mu\nu}}{2}R\right)\nabla_\mu\phi^2, \quad
\nabla_\mu G(T^{\mu\nu}_g+T^{\mu\nu}_n)=0.
\label{DTnDTg}
\end{equation}
We consider here the cosmological constant in the new gravitational equations.
While the scalar equation (\ref{SGEQ}) does not depend on the cosmological constant $\Lambda_0$,  
the tensor equation (\ref{ZEMT}) requires the total cosmological constant to be zero.  
Therefore, $T^{\mu\nu}_n$ should have the cosmological constant $-\Lambda_0g^{\mu\nu}$ 
to cancel the cosmological constant in $T^{\mu\nu}_\phi$, 
since $T^{\mu\nu}_t$ does not have the cosmological constant term by definition.  
This observation shows that new gravity essentially settles the cosmological constant problem \cite{Weinberg}. 

The cosmological constant becomes a serious problem if it originates from the vacuum energy 
or the zero-point energy in quantum field theory, since it is infinite.
In this view, quantum corrections may induce for each constituent energy-momentum tensor:
$T^{\mu\nu}_\phi$, $T^{\mu\nu}_t$, and $T^{\mu\nu}_n$ in (\ref{ZEMT}) the cosmological constant terms of their own, 
$\Lambda_\phi g^{\mu\nu}$, $\Lambda_tg^{\mu\nu}$, and $\Lambda_ng^{\mu\nu}$, respectively. 
Then, the gravitational equation (\ref{ZEMT}) dictates that the total cosmological constant should be zero: 
$\Lambda=\Lambda_\phi+\Lambda_t+\Lambda_n=0$. 
If we adopt the convention that $\Lambda_c=0$, the cosmological constant for the normal matter $\Lambda_n$ should be cancelled by that for gravity $\Lambda_g=\Lambda_s+\Lambda_t$, where $\Lambda_g$ may differ from $\Lambda_0$.
\section{Friedmann equations\label{FEQ}}
We apply new gravitational equations 
(\ref{SGEQ}) and (\ref{EEG}) to the expanding universe and clarify the difference between the Einstein gravity and the new gravity.  
We assume here the Friedmann-Robertson-Walker metric with zero spacial curvature: 
\begin{equation}
ds^2=g_{\mu\nu}dx^\mu dx^\nu=dt^2-a(t)^2\delta_{ij}dx^idx^j, 
\label{FRW}
\end{equation}
and that the energy-momentum tensor for normal matter $T^{\mu\nu}_n$ has the form of a perfect fluid:
\begin{equation}
T^{00}_n=\rho_n, \quad T^{ij}_n=a(t)^{-2}\delta^{ij}p_n. 
\label{TinFRW}
\end{equation}
In this metric, the scalar equation (\ref{SGEQ}) takes the form
\begin{equation}
\ddot{\phi}+3H\dot{\phi}-a^{-2}\partial_i^2\phi-2\phi(\dot{H}+2H^2)+\frac{\partial V(\phi)}{\partial\phi}=0,\quad H=\frac{\dot{a}}{a}, \quad \dot{a}=\frac{da}{dt},
\label{SEQ}
\end{equation}
while the tensor equation (\ref{EEG}) is expressible in the form
\begin{equation}
\left[\begin{array}{c}\rho_c \\p_c  \end{array}\right]
=
\left[\begin{array}{c}\rho_g \\p_g\end{array}\right]
+
\left[\begin{array}{c}\rho_n \\p_n\end{array}\right], 
\label{TEQ}
\end{equation}
where
\begin{eqnarray}
\left[\begin{array}{c}\rho_c \\p_c  \end{array}\right]
&=&
\left[\begin{array}{c}3H^2\phi^2 \\-(2\dot{H}+3H^2)\phi^2 \end{array}\right], \quad 
\left[\begin{array}{c}\rho_g \\p_g\end{array}\right]
:=
\left[\begin{array}{c}\rho_s \\p_s\end{array}\right]
+
\left[\begin{array}{c}\rho_t \\p_t \end{array}\right],
\\
\left[\begin{array}{c}\rho_s \\p_s \end{array}\right]
&=&
\left[\begin{array}{c}
\frac{1}{2}\left(\dot{\phi}-2H\phi\right)^2+V(\phi)\\
\frac{1}{2}\dot{\phi}^2-V(\phi)+\frac{2}{3}\left(\phi\ddot{\phi}+\dot{\phi}^2+2H\phi\dot{\phi}\right) 
-\frac{2}{3}\left(2\dot{H}+3H^2\right)\phi^2
\end{array}\right],
\label{rho_s&p_s}\\
\left[\begin{array}{c}\rho_t \\p_t  \end{array}\right]
&=&
\left[\begin{array}{c}
3g^{-2}\left(H\ddot{H}+3H^2\dot{H}-\frac{1}{2}\dot{H}^2\right)\\
-g^{-2}\left(\dddot{H}+6H\ddot{H}+9H^2\dot{H}+\frac{9}{2}\dot{H}^2\right)
\end{array}\right].
\label{rho_t&p_t}
\end{eqnarray}
\section{Exact solutions\label{ESS}}
If $j_n=0$, the scalar equation (\ref{SEQ}) admits the exact solutions of the form:
\begin{equation}
\phi(t)=\frac{\beta\eta}{\sin\theta}, \quad a(t)=\sin^\gamma\theta,\quad \theta=\alpha t,
\label{EXS}
\end{equation}
where $\alpha$ is a constant of mass dimension one, while $\beta$ and $\gamma$ are dimensionless numbers.
Putting (\ref{EXS}) into (\ref{SEQ}), we find 
\begin{equation}
\alpha^2=\frac{4g^2\eta^2}{3(4\gamma^2+3\gamma-1)}, \quad \beta^2=\frac{4\gamma^2+\gamma-2}{4\gamma^2+3\gamma-1}. 
\label{AB}
\end{equation}
Since these are positive numbers, the value of $\gamma$ is in the range
\begin{equation}
\gamma>\frac{\sqrt{33}-1}{8}=0.593\cdots, \quad {\rm or}
\quad \gamma<-1.
\label{RangeOfGamma}
\end{equation}
The first equation in (\ref{EXS}) implies that $\phi$ starts from the infinity: $\phi(0)=\infty$,  
which is certainly another symmetric point of the potential $V(\phi)$ other than $\phi=0$. 
It will be a natural condition rather than assuming that $\phi$ starts from the top of a small hill of the potential: $V(0)$, 
if the early universe was very hot and dense. 

\begin{figure}[htbp] 
\includegraphics[width=8cm]{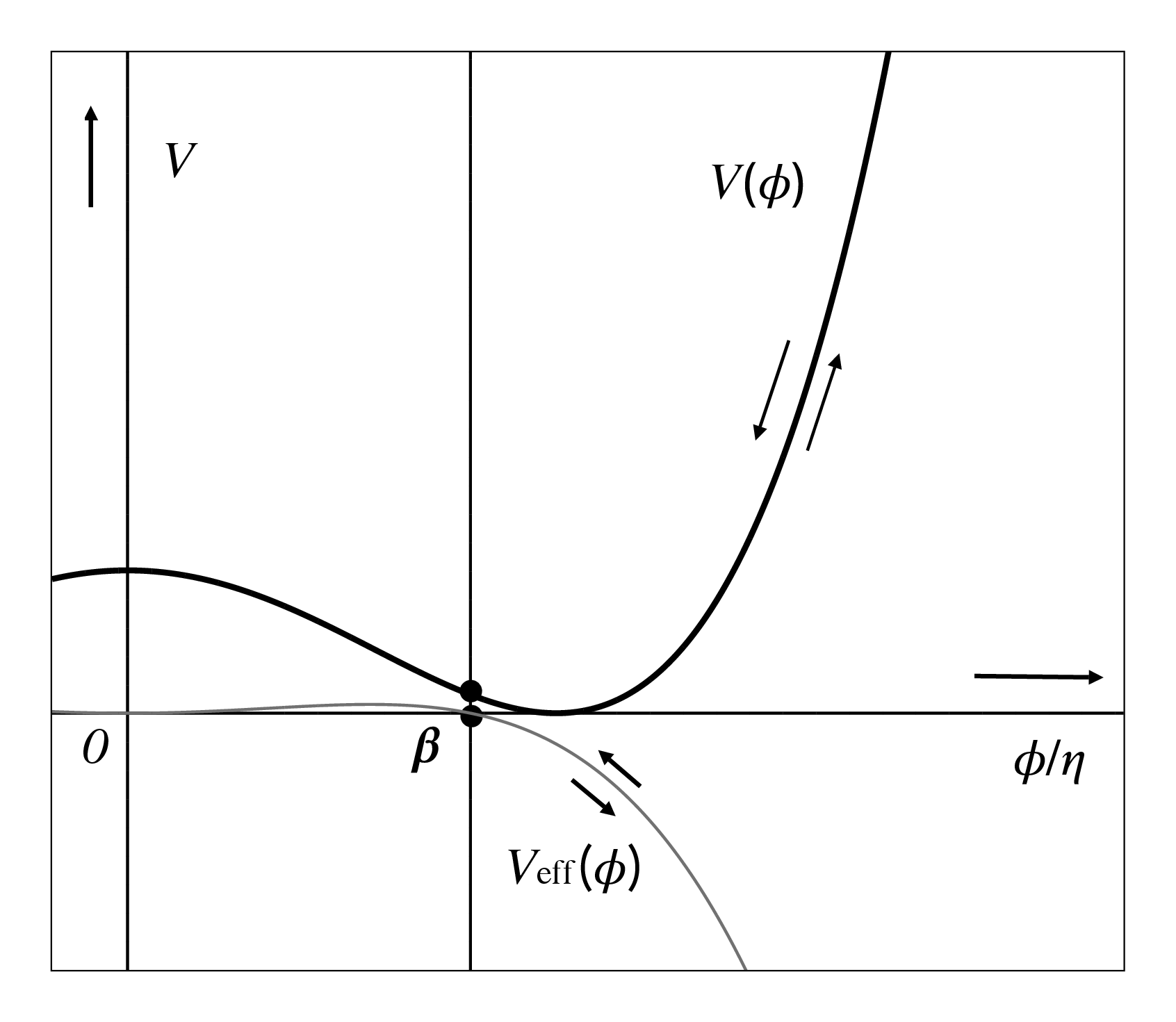}  
\caption{The scalar field $\phi$ in Sec.\ref{ESS}, rolling down and up in the potential $V(\phi)$ (a bold line), while rolling up and down in the effective potential $V_{\rm eff}(\phi)$ (a grey line). The vertical scale is arbitrary, and the slope of $V_{\rm eff}(\phi)$ is exaggerated. The turning point $\phi/\eta=\beta$ corresponds to the maximum of the scale factor $a(t)$.}
\label{Fig_1}
\end{figure}

The energy conservation $\dot{\phi}^2/2+V_{\rm eff}(\phi)=0$ gives the effective potential:  
\begin{equation}
V_{\rm eff}(\phi)=-\frac{2g^2}{3b}\phi^2(\phi^2-\beta^2\eta^2), \quad b:=4\gamma^2+\gamma-2,
\end{equation}
which has a fairly gentle slope for a large $\gamma$. 
In this effective potential, the scalar $\phi$ rolls up from 
$V_{\rm eff}(+\infty)=-\infty$ to $V_{\rm eff}(\beta\eta)=0$, and then returns to the infinity (Fig.\ref{Fig_1}). 

\begin{figure}[htbp] 
\includegraphics[width=11cm]{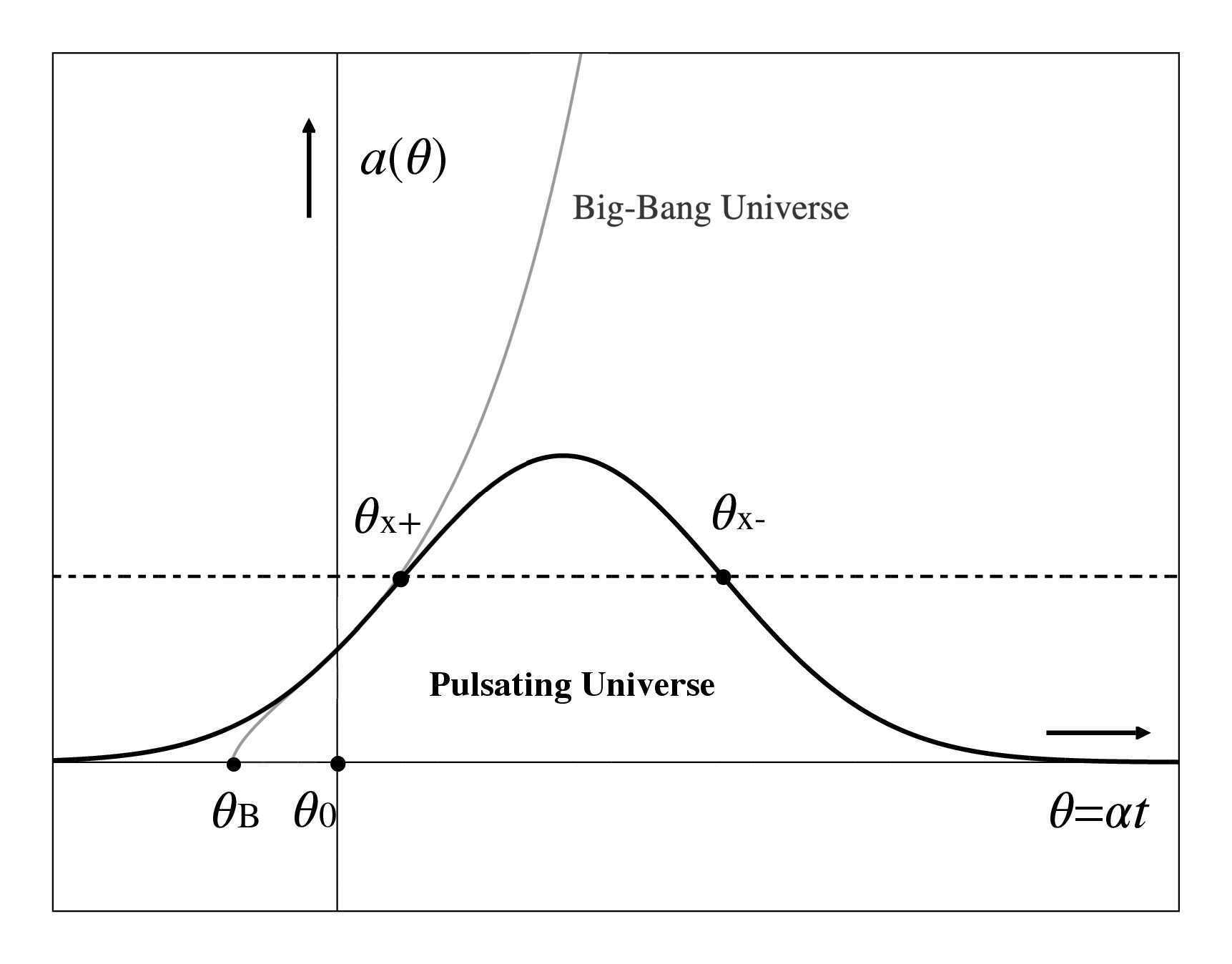} 
\caption{The comparison between the pulsating universe (a bold line) and the Big-Bang universe (a grey thinner line) in Sec.\ref{ESS},  
where $\theta_0$ corresponds to the ``standard present time". The Big-Bang would occur at $\theta_B$, while the pulsating universe experiences the phase transition from inflationary expansion to decelerating expansion at $\theta_{{\rm x}+}$, and the phase transition from accelerating contraction to deflationary contraction at $\theta_{{\rm x}-}$. The portion of the bold line under the horizontal dashed line corresponds to the era of repulsive gravity dominance, while that above to the era of attractive gravity dominance.}
\label{Fig_2}
\end{figure}

The second equation of (\ref{EXS}) implies that the evolution of the universe can be periodic in time, if $\gamma$ is an integer. 
For $\gamma>0$, it begins from a point at $t=0$, and expands to the maximum. 
Then it returns to a point again (Fig.\ref{Fig_2}). 
For $\gamma<0$, on the other hand, it begins from the infinity and contracts to the minimum, 
and then blows up again to the infinity in a finite time.  

The Hubble parameter $H$, and the deceleration parameter $q$ are calculated as
\begin{equation}
H=\alpha\gamma\cot\theta, \quad q=-1-\frac{\dot{H}}{H^2}=-1+\frac{1}{\gamma\cos^2\theta}.
\label{DetOfAlpha}
\end{equation}
From the second equation, we notice that $1+q>0$ for $\gamma>0$, and $1+q<0$ for $\gamma<0$ at all times. 
According to the recent observational data, the present value of the deceleration parameter is $q_0=-0.51$ \cite{Riess}, and therefore $1+q_0>0$. Consequently, the solution with $\gamma<0$ will not correspond to our universe.
We consider in the following only the case with a positive $\gamma$.

Further, if $\gamma>1$ the deceleration parameter $q$ vanishes at the phase transition points 
\begin{equation}
\theta_{\rm x\pm}=\arccos(\pm1/\sqrt{\gamma}).
\label{PTP}
\end{equation}
Then, the universe experiences accelerating expansion during $0<\theta<\theta_{\rm x+}$, 
while decelerating expansion during $\theta_{\rm x+}<\theta<90^\circ$.
From the first equation, we see that $H=0$ at $\theta=90^\circ$, where the universe stops expanding and then begins  contracting to a point again.

There are several reasons to expect that $\gamma$ is large, for example, $\gamma>10$, or $\gamma>100$. 
One reason is that a large $\gamma$ will induce an inflation \cite{Sato,Guth,Lucchin}, 
which seems necessary for understanding the anisotropy of the cosmic microwave background (CMB). 
The second is, as we presently see, that $\gamma$ grows large as the present dark energy component
$\Omega_\Lambda^0$ in the $\Lambda$CDM cosmology approaches $2/3$, 
which approximates the observational value $\Omega_\Lambda^0=0.683$ \cite{PlanckCollaboration}.
Finally, the observational constraint on the time variation of the gravitational constant also requires a large $\gamma$
\cite{Weinberg2}. 

The energy density $\rho_a$ and the pressure $p_a$ in the previous section are expressible in the form:
\begin{eqnarray}
&&\left\{\begin{array}{rcl}
\rho_c=e\left(c_1\sin^{-4}\theta+c_2\sin^{-2}\theta\right),\\
p_c=e\left(c_3\sin^{-4}\theta+c_4\sin^{-2}\theta\right),
\end{array}\right.\label{rho_c}\\
&&\left\{\begin{array}{rcl}
\rho_s&=&e\left(s_1\sin^{-4}\theta+s_2\sin^{-2}\theta+\lambda_s\right),\\
p_s&=&e\left(s_3\sin^{-4}\theta+s_4\sin^{-2}\theta-\lambda_s\right),
\end{array}\right. \\
&&\left\{\begin{array}{rcl}
\rho_t&=&e\left(t_1\sin^{-4}\theta+t_2\sin^{-2}\theta+\lambda_t\right),\\
p_t&=&e\left(t_3\sin^{-4}\theta+t_4\sin^{-2}\theta-\lambda_t\right),
\end{array}\right.\\
&&\left\{\begin{array}{rcl}
\rho_g&=&e\left(g_1\sin^{-4}\theta+g_2\sin^{-2}\theta+\lambda_g\right),\\
p_g&=&e\left(g_3\sin^{-4}\theta+g_4\sin^{-2}\theta-\lambda_g\right),
\end{array}\right.
\label{Rho_gP_g}\\
&&\left\{\begin{array}{rcl}
\rho_n&=&e\left(n_1\sin^{-4}\theta+n_2\sin^{-2}\theta+\lambda_n\right),\\
p_n&=&e\left(n_3\sin^{-4}\theta+n_4\sin^{-2}\theta-\lambda_n\right),
\end{array}\right.
\label{Rho_nP_n}\\
\end{eqnarray}
where we have introduced the partial cosmological constants: $\lambda_s$, $\lambda_t$, $\lambda_g$ and $\lambda_n$, 
in view of the argument in Sec.\ref{GEQ}.  
These can be specified at will under the conditions: $\lambda_g=\lambda_s+\lambda_t$ and $\lambda_g+\lambda_n=0$, 
without affecting the gravitational equations.
The standard cosmological constant $\Lambda_0$ corresponds to $\lambda_s=\lambda_0$, where
\begin{equation}
\lambda_0=\Lambda_0/e=\frac{(b+2\gamma+1)^2}{12b\gamma^2}, \quad e=3(\alpha\beta\gamma\eta)^2.
\end{equation}
The numerical coefficients $c_i$, $s_i$, and $t_i$ are given by
\begin{equation}
\left\{\begin{array}{rcl}
c_1&=&1,\\
c_2&=&-1,\\
c_3&=&\frac{2}{3\gamma}-1,\\
c_4&=&1,
\end{array}\right.
\quad 
\left\{\begin{array}{rcl}
s_1&=&1+\frac{3}{4\gamma},\\
s_2&=&-\frac{4}{3}-\frac{7}{6\gamma},\\
s_3&=&-1-\frac{1}{12\gamma}+\frac{1}{\gamma^2},\\
s_4&=&\frac{4}{3}+\frac{17}{18\gamma}-\frac{7}{9\gamma^2},\\
\end{array}\right.
\quad
\left\{\begin{array}{rcl}
t_1&=&-\frac{4\gamma}{b}\left(1-\frac{1}{2\gamma}\right),\\
t_2&=&\frac{4\gamma}{b}\left(1-\frac{2}{3\gamma}\right),\\
t_3&=&\frac{4\gamma}{b}\left(1-\frac{1}{2\gamma}\right)\left(1-\frac{4}{3\gamma}\right),\\
t_4&=&-\frac{4\gamma}{b}\left(1-\frac{2}{3\gamma}\right)^2,\\
\end{array}\right.
\label{CST}
\end{equation}
while $g_i$ and $n_i$ are given by $g_i:=s_i+t_i$, $n_i=c_i-g_i$, $(i=1,2,3,4)$. 
In particular, the following expressions are useful in later considerations:
\begin{equation}
\left\{\begin{array}{rcl}
n_1&=&-\frac{3}{4\gamma}+\frac{4\gamma}{b}\left(1-\frac{1}{2\gamma}\right)
=\frac{1}{4\gamma}\left(1-\frac{3}{\gamma}+\cdots\right),\\
n_2&=&\frac{1}{3}+\frac{7}{6\gamma}-\frac{4\gamma}{b}\left(1-\frac{2}{3\gamma}\right)
=\frac{1}{3}\left(1+\frac{1}{2\gamma}+\cdots\right),\\
n_3&=&\left(\frac{4}{3\gamma}-1\right)n_1
=-\frac{1}{4\gamma}\left(1-\frac{13}{3\gamma}+\cdots\right),\\
n_4&=&\left(\frac{2}{3\gamma}-1\right)n_2
=-\frac{1}{3}\left(1-\frac{1}{6\gamma}+\cdots\right),\\
\end{array}\right.
\label{UExp}
\end{equation}
where the second equalities are the asymptotic expansions with respect to $\gamma^{-1}$. 

We remark that the critical density $\rho_c=e\sin^{-4}\theta\cos^2\theta$ becomes zero when the universe stops expanding at $\theta=90^\circ$. 
Accordingly, it will be convenient to calibrate the energy densities $\rho_s$, $\rho_t$, $\rho_g$, and $\rho_n$ so that each component also becomes zero at $\theta=90^\circ$. This is allowable, and realizes if we take 
$\lambda_a=-a_1-a_2$ for $a=s, t, g, n$. In this case, $\lambda_g$ or $-\lambda_n$ is found very close to $\lambda_0$ for a large $\gamma$:
\begin{equation}
\lambda_g=-\lambda_n=\lambda_0+\frac{1}{12\gamma^2}+{\cal O}(\gamma^{-3}).
\end{equation}
\section{Dark energy and Dark matter \label{DEDM}}
We first introduce the notion of the ``cosmological gauge transformation" 
for the energy momentum tensor $T^{\mu\nu}_a$:
\begin{equation}
T^{\mu\nu}_a\rightarrow T^{\mu\nu}_a+\Lambda_a(t)g^{\mu\nu}, \quad {\rm or}, \quad 
\left[\begin{array}{c}
\rho_a \\
p_a   
\end{array}\right]
\rightarrow
\left[
\begin{array}{c}
\rho_a+\Lambda_a(t) \\
p_s-\Lambda_a(t)   
\end{array}\right], \quad (a=s, t, g, n),
\label{CGT}
\end{equation}
where the gauge parameter $\Lambda_a(t)$ is a function of the cosmological time $t$.  
When $\Lambda_a(t)$ is constant, it becomes an ordinary cosmological constant.  
We do not consider the gauge transformation of the critical density $ \rho_c$, nor the critical pressure $p_c$, 
since these are regarded as fiducial quantities. 

We consider again the cosmological constant problem \cite{Weinberg} in the context of our new formalism. 
If the gravitational equation is invariant under the above gauge transformation, then the cosmological constant problem will disappear, since the problem becomes simply a matter of gauge fixing.

While the scalar equation (\ref{SEQ}) is invariant under any change of gauge parameter $\Lambda_s$, 
the tensor equation (\ref{ZEMT}), or (\ref{TEQ}) is invariant under the condition 
$\Lambda_g+\Lambda_n=\Lambda_s+\Lambda_t+\Lambda_n=0$. 
However, as is well known, since the Friedmann equations for the energy density component and the pressure component (\ref{TEQ}) are equivalent, the two equations can be cast into the gauge invariant form in terms of the ``invariant densities" $\tilde{\rho}_a:=\rho_a+p_a$: 
\begin{equation}
\tilde{\rho}_c=\tilde{\rho}_g+\tilde{\rho}_n, \quad{\rm or}, \quad 1=\tilde{\Omega}_g+\tilde{\Omega}_n, \quad 
\tilde{\Omega}_g:=\frac{\tilde{\rho}_g}{\tilde{\rho}_c},\quad \tilde{\Omega}_n:=\frac{\tilde{\rho}_n}{\tilde{\rho}_c},
\label{IES}
\end{equation}
which contrasts to the ordinary (gauge dependent) energy spectrum of the universe:
\begin{equation}
\rho_c=\rho_g+\rho_n, \quad{\rm or}, \quad 1=\Omega_g+\Omega_n, \quad 
\Omega_g:=\frac{\rho_g}{\rho_c},\quad \Omega_n:=\frac{\rho_n}{\rho_c}.
\label{OES}
\end{equation}
Consequently, the equation (\ref{IES}) for the tensor gravity also becomes completely independent of the gauge parameters. 

We next consider the physical implication of the invariant densities $\tilde{\rho}_a$. 
Since the new gravitational equations  (\ref{SEQ}) and (\ref{IES}) are ``gauge invariant", 
we can take use of this gauge freedom to convert each specific energy-momentum tensor into the form of ``cold matter" with zero pressure:
\begin{equation}
\left[\begin{array}{c}
\rho_a \\
p_a   
\end{array}\right]
\rightarrow
\left[
\begin{array}{c}
 \rho_a+\Lambda_a(t)\\
p_a -\Lambda_a(t)   
\end{array}\right]
=
\left[
\begin{array}{c}
 \rho_a+p_a\\
0  
\end{array}\right], 
\quad \Lambda_a(t)=p_a, \quad (a=g,n).
\label{CMTr}
\end{equation}
We find from (\ref{CMTr}) that the invariant density of $a$-fluid: $\tilde{\rho}_a=\rho_a+p_a$ is the effective energy density as cold matter. 
Then, the Friedman equation in the gauge invariant form (\ref{IES}) implies the energy spectrum of the cold matter. 

In order to find out the correspondence of the quantities $\Omega_g$, $\Omega_n$, $\tilde{\Omega}_g$, and $\tilde{\Omega}_n$ to those in the Big-Bang cosmology, 
we cast the Friedmann equations (\ref{TEQ}) into the $\Lambda$CDM spectrum: 
\begin{equation}
\left[\begin{array}{c}
\rho_c \\
p_c   
\end{array}\right]
=
\left[
\begin{array}{c}
 \Lambda_d\\
-\Lambda_d   
\end{array}\right]
+
\left[
\begin{array}{c}
 \rho_m\\
0  
\end{array}\right], 
\label{LambdaCDM}
\end{equation}
where $\Lambda_d$ is the cosmological constant as the ``dark energy", and $\rho_m$ is the cold matter density. 
Then, we find the expressions of $\Lambda_d$ and $\rho_m$ in the new formalism:
\begin{equation}
\Lambda_d=-p_c, \quad  \rho_m=\rho_c+p_c=\tilde{\rho}_c,  
\label{LambdaRho_m}
\end{equation}
from which we have 
\begin{equation}
\Omega_\Lambda:=\frac{\Lambda_d}{\rho_c}=1-\Omega_m,\quad
\Omega_m:=\frac{\rho_m}{\rho_c}=\frac{\tilde{\rho}_c}{\rho_c}=\Omega_{\rm gcm}+\Omega_{\rm ncm}, \quad
\Omega_{\rm gcm}:=\frac{\tilde{\rho}_g}{\rho_c}=\Omega_m\tilde{\Omega}_g,\quad
\Omega_{\rm ncm}:=\frac{\tilde{\rho}_n}{\rho_c}=\Omega_m\tilde{\Omega}_n,
\label{Omega_Lm}
\end{equation}
where $\Omega_{\rm gcm}$ is the component for the gravitational cold matter, 
while $\Omega_{\rm ncm}$ is that for the normal cold matter.
Therefore, the following identifications are natural:
\begin{equation}
\Omega_d=\Omega_{\rm gcm}, \quad \Omega_b=\Omega_{\rm ncm},
\label{CDMB}
\end{equation}
where $\Omega_d$ is for the ``cold dark matter", while  $\Omega_b$ is for the (cold) baryonic matter,  
since the dominant component of the normal cold matter will be baryons. 

We next examine the correspondences (\ref{Omega_Lm}) and (\ref{CDMB}) quantitatively, in particular, for the present values. 
In order to do that, it is convenient to specify the ``standard present time" $t_0$ by $q_0=-1/2$.
Since
\begin{equation}
\Omega_\Lambda=-\frac{p_c}{\rho_c}=1+\frac{2\dot{H}}{3H^2}=\frac{1-2q}{3},\quad 
\Omega_m=\frac{\tilde{\rho}_c}{\rho_c}=\frac{2}{3\gamma\cos^2\theta}, 
\end{equation}
we obtain the relations:   
\begin{equation}
\cos\theta_0=\sqrt{\frac{2}{\gamma}}, \quad \theta=\alpha t=\frac{\sqrt{\frac{\gamma}{2}-1}}{\gamma}H_0t, \quad
t_0=\frac{\gamma }{\sqrt{\gamma/2-1}}H_0^{-1}\arccos\sqrt{\frac{2}{\gamma}},
\label{StandardPresentTime}
\end{equation}
where $H_0$ is the Hubble constant at the standard present time. 
Then, we have by definition $(\Omega_\Lambda^0, \Omega_m^0)=(2/3,1/3)$, which approximate well those obtained by the astronomical observations: $\Omega_\Lambda^0=0.683$ \cite{PlanckCollaboration}, and $q_0=-0.51$ \cite{Riess}. 
Owing to (\ref{UExp}), we find the expressions 
\begin{eqnarray}
\Omega_n&=&1-\Omega_g=n_1+(n_1+n_2)\sin^2\theta,\\
\tilde{\Omega}_n&=&1-\tilde{\Omega}_g=2n_1+n_2\sin^2\theta,  
\end{eqnarray}
where $\Omega_n$ and $\tilde{\Omega}_n$ increase as the universe expands, and decrease as the universe contracts. These normal matter components, or the gravitational components, take nearly the same value for a large $\gamma$:
\begin{equation}
\tilde{\Omega}_n-\Omega_n=\Omega_g-\tilde{\Omega}_g=n_1\cos^2\theta\simeq\frac{\cos^2\theta}{4\gamma}.  
\end{equation}
Further, we obtain as the present values 
\begin{eqnarray}
\Omega_n^0&=&1-\Omega_g^0=\frac{1}{3}-\frac{17}{12\gamma^2}+\cdots,\\
\tilde{\Omega}_n^0&=&1-\tilde{\Omega}_g^0=\frac{1}{3}-\frac{11}{12\gamma^2}+\cdots, 
\end{eqnarray}
and therefore,  
\begin{equation}
\Omega_g^0\simeq2/3,\quad \Omega_d^0\simeq2/9,\quad \Omega_b^0\simeq1/9,
\label{Omegas}
\end{equation}
which are compared to the observational values: 
$(\Omega_\Lambda^0,\Omega_d^0,\Omega_b^0)=(0.683,0.268,0.049)$ \cite{PlanckCollaboration}. 
Then, we confirm from $(\Omega_g^0,\Omega_n^0)\simeq(\Omega_\Lambda^0, \Omega_m^0)$ that 
the true nature of dark energy is the gravitational energy,  
though they behave differently as the universe expands, 
since the dark energy $\Lambda_d$ and the gravitational energy $\rho_g$ obey different equations of state. 
The same caution applies for the cold matter $\rho_m$ in the $\Lambda$CDM cosmology and the normal matter $\rho_n$ in the new cosmology.

\begin{figure}[htbp] 
\includegraphics[width=11cm]{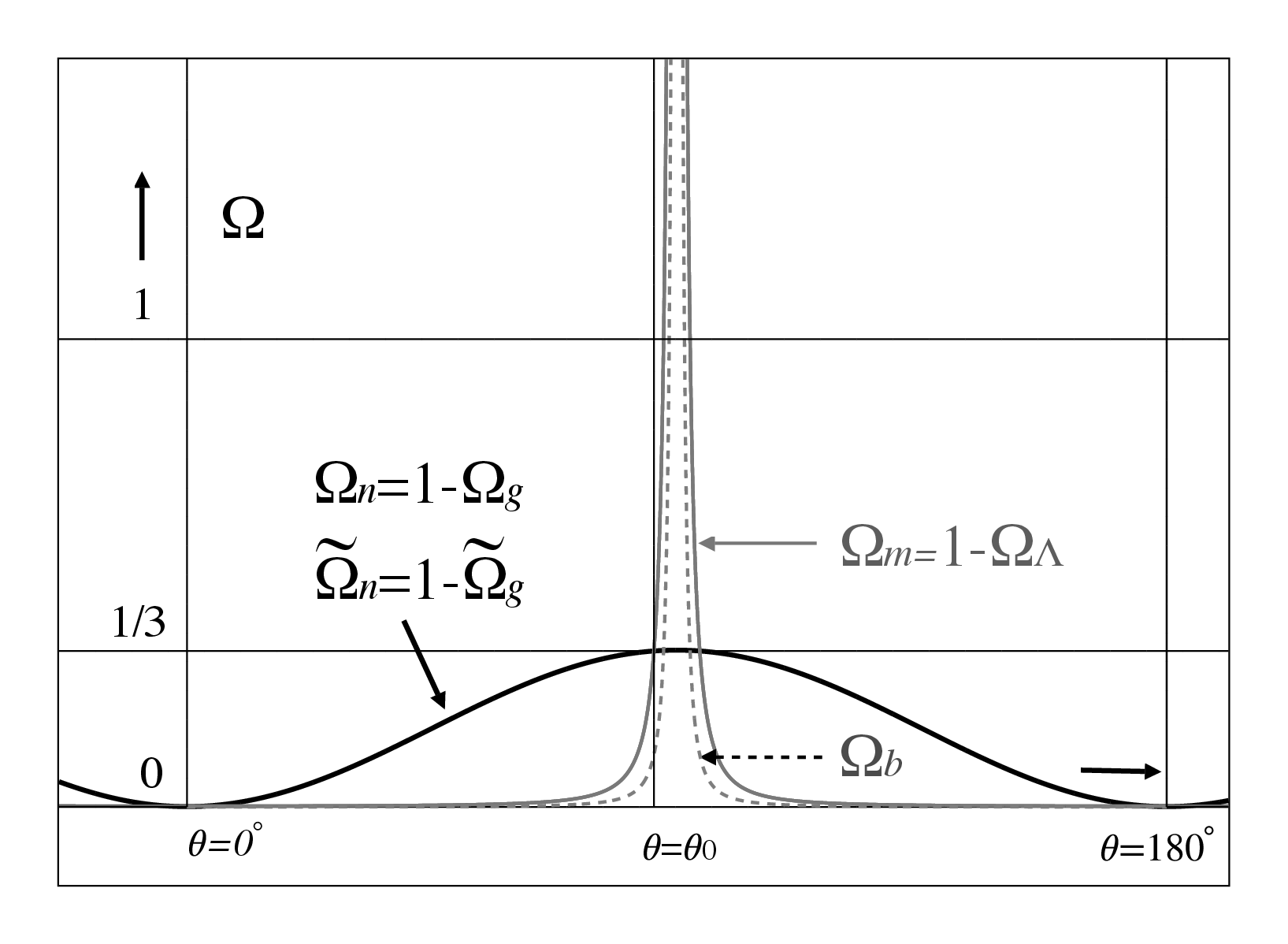} 
\caption{The time evolution of the energy spectrum in the pulsating universe in Sec.\ref{DEDM}, 
and that casted into the $\Lambda$CDM spectrum. The normal matter component $\Omega_n(\theta)$ 
and the normal cold matter component $\tilde{\Omega}_n(\theta)$ (bold lines) are regular functions of $\theta$, though graphically not distinguishable, 
while $\Omega_m=1-\Omega_\Lambda$ (a grey line) is singular at $\theta=90^\circ$. 
Nevertheless, the graph shows that $\Omega_m$ and $\Omega_n$ take almost the same value at the standard present time $t_0=\alpha^{-1}\theta_0$.
The region sandwiched between $\Omega_m$ and $\Omega_b$ corresponds to the cold dark matter $\Omega_d$, though it is the mirage of the distinction between the gravitational energy and the normal matter energy in the new cosmology.}
\label{Fig_3}
\end{figure}

We find further that the distinction between the cold dark matter and the baryons (normal cold matter) 
in the $\Lambda$CDM cosmology is simply the reflection of the distinction between the gravitational energy and the normal matter energy 
caused by casting them into the $\Lambda$CDM spectrum (Fig.\ref{Fig_3}).  

The value of $\Omega_b^0$ in (\ref{Omegas}) is a little large compared to the observational data, 
possibly since we assumed tentatively that the normal matter consists only of baryons.
\section{Varying equation of state\label{VES}}
When $\gamma$ is large, the pulsating universe (\ref{EXS}) begins with inflation. 
The era of inflation continues till the universe reaches the phase transition point $\theta=\theta_{\rm x+}$ (\ref{PTP}), 
and then enters the era of decelerating expansion.
Astronomical observations reveal that our universe is still in the phase of accelerating expansion \cite{Riess0,Riess,Perlmutter}, and therefore $\theta_0<\theta_{\rm x+}$. 
The cause of the phase transition at $\theta_{\rm x+}$ in the pulsating universe is inferred from the equations of state for 
the normal matter and gravitational fluids: $p_a=w_a\rho_a, \quad (a=n,g)$.

We first consider the case of normal matter $w_n=p_n/\rho_n$ (\ref{Rho_nP_n}). 
Owing to the relations (\ref{UExp}), we have
\begin{equation}
w_n(\theta)=-1+\frac{2}{3\gamma\cos^2\theta}\left[1+\frac{n_1\cos^2\theta}{n_1+(n_1+n_2)\sin^2\theta}\right].
\label{w_n}
\end{equation}
Then, $w_n$ increases monotonically from a negative value $w_n(0^\circ)$ to the positive infinity $w_n(90^\circ)=+\infty$,  
where
\begin{equation}
w_n(0^\circ)=-1+\frac{4}{3\gamma}.
\end{equation}
Accordingly, the normal matter fluid behaves like a cosmological constant (the dark energy) $w=-1$ at the beginning of the universe, and then becomes to behave successively like a cold matter, a radiation, and finally 
an ideal gas $w=\infty$ (since $\rho\propto T^{1+1/w}$ for a perfect fluid) at $\theta=90^\circ$. 
In particular, when the universe is at the phase transition point $\theta_{\rm x+}$, $w_n$ takes the value
\begin{equation}
w_n(\theta_{\rm x+})=-\frac{1}{3}+{\cal O}(\gamma^{-2}).
\end{equation}

\begin{figure}[htbp] 
\includegraphics[width=10cm]{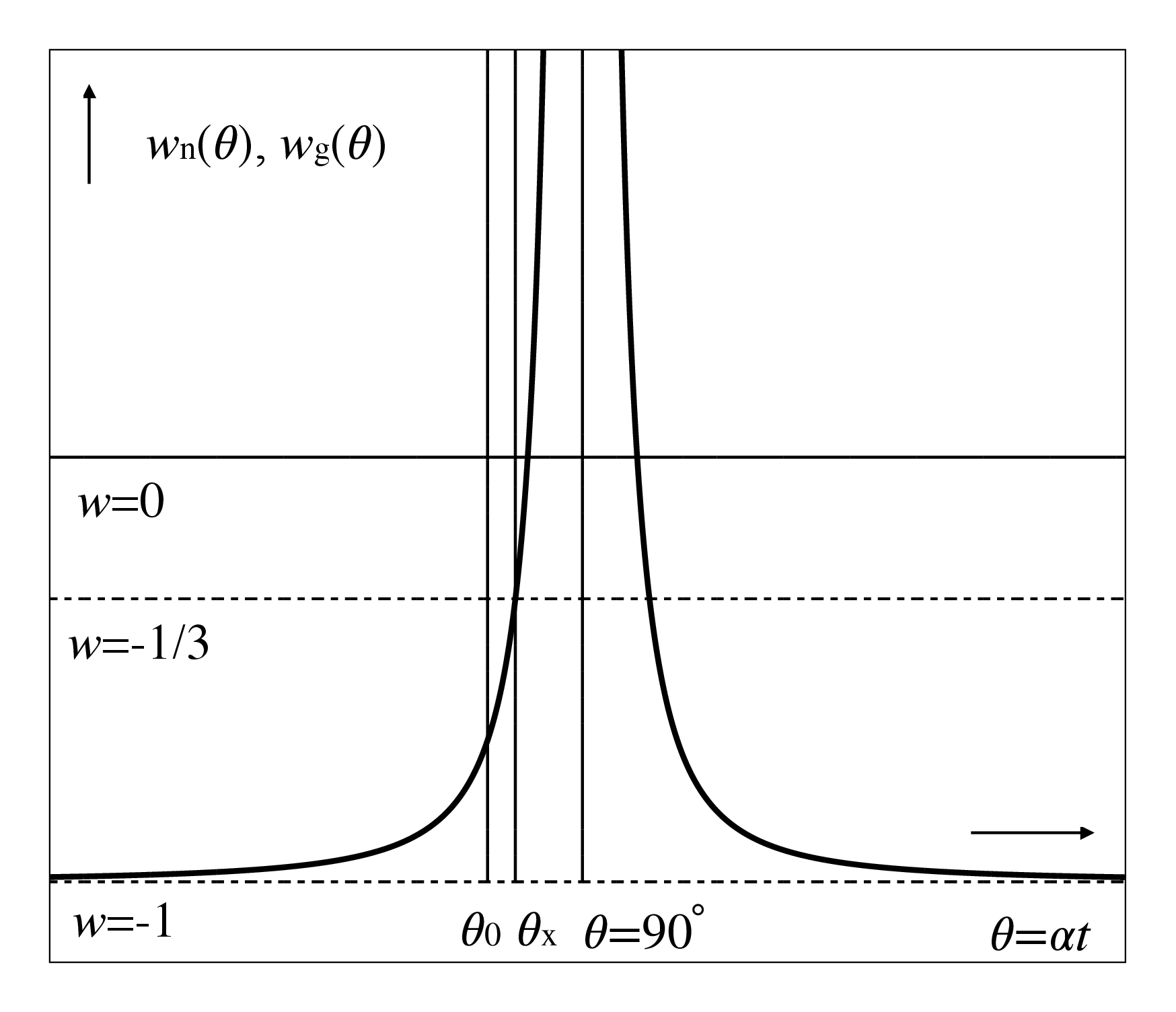} 
\caption{The varying equations of state: $w_n=p_n/\rho_n$, and $w_g=p_g/\rho_g$ in Sec.\ref{VES}. The both are graphically not distinguishable. We see that $w\simeq-1/3$ at the phase transition point $\theta=\theta_{\rm x+}$, 
where the era of repulsive gravity ends and the era of attractive gravity begins. We are at present still in the era of repulsive gravity: $w_n^0<-1/3$ and $w_g^0<-1/3$.}
\label{Fig_4}
\end{figure}

On the other hand, we have for the gravitational fluid $w_g=p_g/\rho_g$ (\ref{Rho_gP_g}): 
\begin{equation}
w_g(\theta)=-1+\frac{2}{3\gamma\cos^2\theta}\left[1-\frac{n_1\cos^2\theta}{1-n_1-(n_1+n_2)\sin^2\theta}\right], 
\end{equation}
which shows that $w_g$ increases monotonically from a negative value $w_g(0^\circ)$ to the positive infinity $w_g(90^\circ)=+\infty$, where
\begin{equation}
w_g(0^\circ)=-1+\frac{2}{3\gamma}+{\cal O}(\gamma^{-2}). 
\end{equation}
We have also in this case,
\begin{equation}
w_g(\theta_{\rm x+})=-\frac{1}{3}+{\cal O}(\gamma^{-2}),
\end{equation}
which is approximately the same as $w_n(\theta_{\rm x+})$, if $\gamma$ is large.   
We see that the two fluids obey almost the same equation of state $p=-\rho/3$ at the phase transition point 
$\theta_{\rm x+}$, which looks like a kind of thermal radiation with the opposite pressure (Fig.\ref{Fig_4}).

We next show that the appearance of this ``thermal anti-radiation" is the symptom that the gravitational force changes its sign. 
The linear gravity approximation gives as the gravitational potential 
$V_{\rm G}$ between small portions of the static fluid with $w_1=p_1/\rho_1$ at $\bm{x}=\bm{x}_1$, 
and that with $w_2=p_2/\rho_2$ at $\bm{x}=\bm{x}_2$:
\begin{eqnarray}
V_{\rm G}&=&-G\int \frac{d^3x_1d^3x_2}{\vert\bm{x}_1-\bm{x}_2\vert}
\left[T^{\mu\nu}(\bm{x}_1)T_{\mu\nu}(\bm{x}_2)-\frac{1}{2}T^\rho{}_\rho(\bm{x}_1)T^\sigma{}_\sigma(\bm{x}_2)\right]\nonumber\\
&=&-G\zeta(w_1,w_2)\frac{m_1m_2}{r_{12}},\quad
\zeta(w_1,w_2):=1+3w_1+3w_2-3w_1w_2.
\end{eqnarray}
where $m_i:=\rho_id^3x_i$. 
The gravitational force becomes repulsive for $\zeta(w_1,w_2)<0$. 
In the case of two portions of the normal matter fluid: $w_1=w_2=w_n$, for example, we have
\begin{equation}
\zeta(w_n,w_n)=1+6w_n-3w_n^2<0 \quad {\rm for}\quad w_n<1-\frac{2}{\sqrt{3}}=-0.1547\cdots, \quad 
w_n>1+\frac{2}{\sqrt{3}}=2.1547\cdots.
\end{equation}
If one of the fluids is a cold matter: $w_2=0$, we have instead,
\begin{equation}
\zeta(w_n,0)=1+3w_n<0 \quad {\rm for}\quad w_n<-\frac{1}{3}.
\end{equation}
In either case, the normal matter exerts repulsive force when $w_n<-1/3$.
If one of the fluids is the thermal radiation $w=1/3$, on the other hand, the gravity returns to attractive: 
\begin{equation}
\zeta(w_n,1/3)=2+2w_n>0 \quad {\rm for}\quad w_n>-1.
\end{equation}
Accordingly, the phase transition point $\theta_{\rm x+}$ corresponds to the time at which 
$w_n$ and $w_g$ begin to exceed $-1/3$, and therefore, at which the both fluids cease to repel, and begin to attract the cold matter.
In this view, the cause of the accelerating expansion of the universe is attributed to the effect of repulsive gravity, and the subsequent decelerating expansion is due to the change of the nature of gravity from repulsive into attractive. 
\section{Thermal property of the universe\label{CMB}}
The Big-Bang cosmology assumes that the cosmic microwave background (CMB) at present originates from the adiabatic cooling of radiation, which was decoupled from thermal equilibrium with matter at redshift $z\simeq10^3$. 
This hypothesis is, however, no longer maintainable in the pulsating universe. 
In fact, if the adiabatic cooling of thermal radiation is assumed, 
we have the relation for the energy density of radiation $\rho_R$: 
\begin{equation}
\frac{\rho_R}{\rho_R^0}=(1+z)^4,
\end{equation}
while the critical density $\rho_c$ (\ref{rho_c}) behaves as
\begin{equation}
\frac{\rho_c}{\rho_c^0}=\frac{\gamma}{2}\left[(1+z)^{4/\gamma}-\left(1-\frac{2}{\gamma}\right)(1+z)^{2/\gamma}\right].
\end{equation}
Then, if $\Omega_R^0=\rho_R^0/\rho_c^0=5\times10^{-5}$ for $T_0=2.725$K is assumed, 
$\rho_R$ would surpass $\rho_c$ at $z\simeq21$ for $\gamma=10$, or  at $z\simeq16$ for $\gamma=100$, 
which implies that the actual temperature of CMB at $z\simeq10^3$ was rather low, and it has experienced reheating after the decoupling from matter at $T\simeq3000$K. 

The energy density of radiation $\rho_R$ may be calculated by recasting the Friedmann equation (\ref{TEQ}) into the sum of the cosmological parameter, the cold matter, and the thermal radiation:
\begin{equation}
\left[\begin{array}{c}
\rho_c \\
p_c   
\end{array}\right]
=
\left[
\begin{array}{c}
 \Lambda_d\\
-\Lambda_d   
\end{array}\right]
+
\left[
\begin{array}{c}
 \rho_m\\
0  
\end{array}\right] 
+
\left[
\begin{array}{c}
 \rho_R\\
\rho_R/3 
\end{array}\right], 
\label{LambdaMR}
\end{equation}
from which we have
\begin{equation}
\rho_m=(1-r)\tilde{\rho}_c, \quad \tilde{\rho}_R=\frac{4}{3}\rho_R=r\tilde{\rho}_c, \quad (0\leq r \leq1). 
\label{R_ratio}
\end{equation}
The ratio $r$ is determined as follows.
We may assume that $r(t=0)=1$ at the beginning of the universe, 
since all the known elementary particles, including the scalar graviton and the tensor graviton, would be in thermal equilibrium with the photon. 
Due to the periodical nature of time evolution of the universe, $r$ will be expressible as the sum of the Legendre polynomials, or the power series of $\cos\theta$. Accordingly, we may assume as the most dominant contribution the simplest form  $r=\cos^\nu\theta$, where the number $\nu$ is not necessarily an integer. 
Then, we have
\begin{equation}
\rho_R=\frac{e}{2\gamma}\sin^{-4}\theta\cos^\nu\theta
\label{Rho_R}
\end{equation}
from which $\cos^\nu\theta_0=4\Omega_R^0=2\times10^{-4}$ follows. 
We next define the ``reduced temperature" $T_R$ by 
\begin{equation}
\rho_R=\frac{\pi^2}{15}T_R^4, \quad T_R:=N_\gamma^{1/4}T, 
\label{TTR}
\end{equation}
where $N_\gamma$ ($1\leq N^{1/4}_\gamma<3$) is the effective multiplicity of the photon, which is half the total degrees of freedom of particles in thermal equilibrium with the photon. 
Then, we have
\begin{equation}
\frac{T_R}{T_0}=\frac{\sin\theta_0}{\sin\theta}\left(\frac{\cos\theta}{\cos\theta_0}\right)^{\nu/4}
\label{T_R}
\end{equation}
We can approximate the formula (\ref{T_R}) by
\begin{equation}
\frac{T_R}{T_0}=(4\Omega_R^0)^{-1/4}\frac{\sin\theta_0}{\sin\theta}
=(4\Omega_R^0)^{-1/4}(1+z)^{1/\gamma},\quad 
\left(\sin\theta<\sin\theta_\epsilon:=\sqrt{\frac{\epsilon\ln(\gamma/2)}{-\ln(4\Omega_R^0)}}\right),
\label{TR_Approx}
\end{equation}
within the error $\epsilon=1-\cos^{\nu/4}\theta$. 
For CMB,  we have $\sin\theta_{\rm cmb}\simeq7.64\times10^{-3}$ with  
$\sin\theta_\epsilon\simeq4.35\times10^{-2}$ for $\epsilon=10^{-2}$ and $\gamma=10$.
Then, the formula (\ref{TR_Approx}) is correct within the error of $1\%$ for $T_R>3000$K and $\gamma>10$,   
and we find that the redshift at $T=T_R=3000$K is $z\simeq10^{2.1\gamma}$ in the new cosmology.
 
Further, when $\theta<\theta_{\rm cmb}$, $\sin\theta$ is well approximated by $\alpha t$, 
and the formula (\ref{TR_Approx}) is further reduced to the form 
\begin{equation}
t=(\Omega_R^0)^{-1/4}H_0^{-1}\sqrt{\gamma}\frac{T_0}{T_R}=43.4\sqrt{\gamma}{\rm y}\left[\frac{10^{10}{\rm K}}{T_R}\right],
\label{t-T_relation}
\end{equation}
where the second equality is for $H_0=73$Kms$^{-1}$Mpc$^{-1}=(13.4{\rm Gy})^{-1}$.  
We find that $t_{\rm cmb}>0.46$Gy for $\gamma>10$, 
which is compared to $t_{\rm cmb}\simeq0.38$My in the Big-Bang cosmology. 

If CMB at the period of $T=T_R\simeq3000$K was cooled down by adiabatic expansion, then the present temperature of CMB would be less than $3000/(1+z)\simeq3\times10^{-18}$K despite our assumption $T_0=2.725$K. 
This result confirms that CMB has been reheated after the decoupling. 
In fact, we have from (\ref{Rho_R})
\begin{equation}
\frac{\dot{\rho}_R}{\rho_R}+4H=\alpha[4(\gamma-1)\cot\theta-\nu\tan\theta]\neq0.
\end{equation}
Since $\alpha\propto H_0$, however, the total heat added during $\Delta t$ is proportional to $H_0\Delta t$,  
and the adiabatic hypothesis is still well applied for the processes accomplished within a very short time relative to the age of the universe. 
The similar argument applies also for the cosmological nucleosynthesis \cite{ABG,Steigman} and baryogenesis \cite{Sakhalov,Trodden}. 

The period of Big-Bang nucleosynthesis (BBN) \cite{ABG} ends when the free neutrons decay away, which is characterized by the temperature $T=1.5\times10^{10}{\rm K}\simeq1.29{\rm MeV}$, 
or  $T_R=N_\gamma^{1/4}T=2.3\times10^{10}{\rm K}$ for $N_\gamma=43/8$ \cite{Weinberg3}.
Then, in the new cosmology, the time available for the cosmological nucleosynthesis will be 
$t_{\rm cns}=19\sqrt{\gamma}>60$y for $\gamma>10$, and accordingly all the chemical elements in the universe could be synthesized by that time,  
whereas in the Big-Bang cosmology only a few light elements can be synthesized within a few minutes. 

However, the formula (\ref{TR_Approx}) shows that the redshift corresponding to that time is 
$z_{\rm cns}\simeq10^{9\gamma}$, while the normal cold matter density $\tilde{\rho}_n$ at that time is
\begin{equation}
\tilde{\rho}_n\simeq\frac{\rho_c^0}{6\gamma}(1+z_{\rm cns})^{4/\gamma}.
\end{equation}
If $\tilde{\rho}_n$ is identified with the baryonic energy density, the contribution of the cosmological nucleosynthesis to the present baryon density will be 
\begin{equation}
\Delta\Omega_b^0=\frac{1}{6\gamma}(1+z_{\rm cns})^{4/\gamma-3}<10^{-9(3\gamma-4)}.
\end{equation}
We further imagine the epoch when the entire universe was filled with nucleons.   
Due to the saturation property of nucleons, the energy density at that time would be $(180{\rm MeV})^4$. 
If this is identified with $\tilde{\rho}_n$ at that time, we have 
\begin{equation}
1+z_{\rm nuc}=\left[6\gamma\frac{\tilde{\rho}_n}{\rho_c^0}\right]^{\gamma/4}
>\left[\frac{\tilde{\rho}_n}{\rho_c^0}\right]^{\gamma/4}\simeq10^{11\gamma}.
\end{equation}
Then, the contribution of baryons at that time to the present baryon density would be
\begin{equation}
\Delta\Omega_b^0=\frac{1}{6\gamma}(1+z_{\rm nuc})^{4/\gamma-3}<10^{-11(3\gamma-4)}.
\end{equation}
Both cases show that these contributions are completely negligible. 
Nevertheless, we have already seen that the present proportion of baryons is roughly 10$\%$ in the critical density (\ref{Omegas}). 
Consequently, either case shows that most of the baryons in the present universe have been (re-)generated after the cosmological (Big-Bang) nucleosynthesis.
How this can be realized is inferred from the following observation. 

The intrinsic state of baryons in the pulsating universe is the normal matter satisfying the equation of state (\ref{w_n}), where the cold baryons are related to the normal matter by the gauge transformation. 
We have seen in Sec.\ref{VES} that 
the state of normal matter changes from the dark energy like state: $w_n\simeq-1$ 
to the cold matter like state: $w_n\simeq0$ as the universe expands.  
When the normal matter is in the ``dark state": $-1<w_n<0$, 
the photons and the baryons included: the ``dark photons" and the ``dark baryons"  
will have the tendency to proliferate in the expanding universe.

In fact, the normal matter satisfies the conservation law (\ref{DTnDTg}), and therefore experiences adiabatic expansion. For a short time during which $w_n$ can be regarded as constant, 
the thermodynamics gives the relations for an adiabatic process: 
\begin{equation}
\rho_n\propto V^{-1-w_n}, \quad U_n=\rho_nV\propto V^{-w_n},
\end{equation}
where $V$ is some volume, for which we may take the intrinsic volume $a^3$. 
When the normal matter, including the photon gas and the baryonic fluid, is in the dark state: $-1<w_n<0$, $\rho_n$ diverges, while the internal energy $U_n$ vanishes as the universe shrinks to a point. 
Then, there would be no photons nor baryons at the beginning of the universe, 
which implies that the universe would be too small to accommodate even a single elementary particle. 
Contrary, $\rho_n$ decreases, while $U_n$ increases as the universe expands, which will explain the reheating of the thermal radiation and the increase of the baryon number in the expanding universe. 

The above observation suggests further that the temperature of the universe at the beginning would be very low, contrary to the Big-Bang hypothesis, 
which can be confirmed by considering the temperature of the normal matter itself.

\begin{figure}[htbp]
\includegraphics[width=10cm]{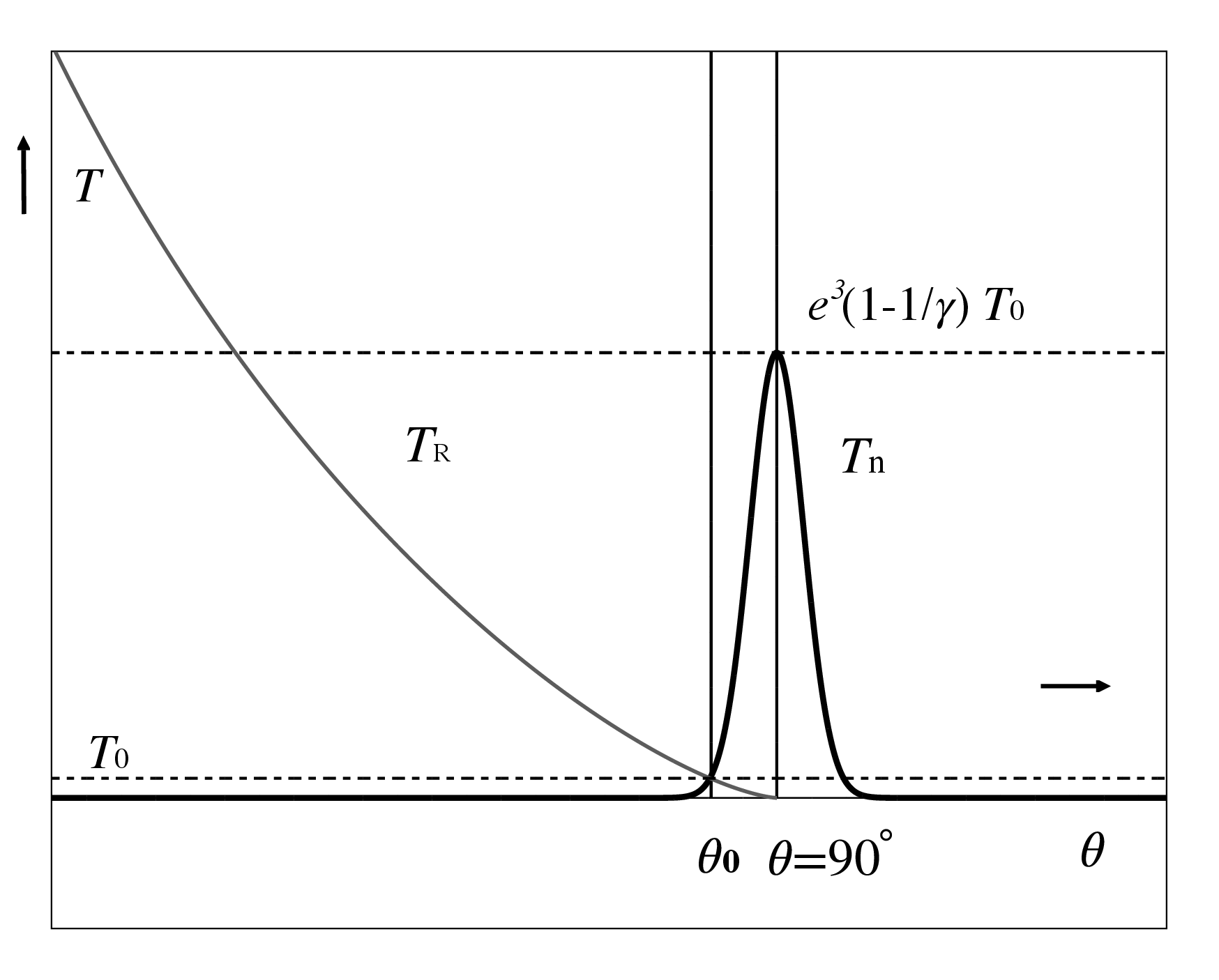} 
\caption{The time evolution of the reduced temperature of thermal radiation $T_R$ and that of normal matter $T_n$ in Sec.\ref{CMB}, when the both are the same at the standard present time. $T_n$ is almost zero for $\theta<\theta_0$. 
The maximum value of $T_n$ at $\theta=90^\circ$ is $e^3(1-1/\gamma)T_0$. }
\label{Fig_5}
\end{figure}

Actually, if we assume $j_n=0$ and the adiabatic expansion for the normal matter, 
the equation of thermodynamics gives 
\begin{equation}
dS=\frac{\rho'_nV}{T_n}dT_n+\frac{\rho_n+p_n}{T_n}dV,
\end{equation}
where the prime implies the derivative with respect to the temperature $T_n$.
The integrability condition for the entropy gives
\begin{equation}
\frac{dT_n}{T_n}=\frac{dp_n}{\rho_n+p_n}=-\frac{ds^2}{s^2}\frac{2n_3+n_4s^2}{n_1+n_3+(n_2+n_4)s^2},
\end{equation}
where $s=\sin\theta$.
Using the relations (\ref{UExp}), we can integrate it to obtain  
\begin{equation}
T_n\propto\left(\frac{2n_1}{n_2}+\sin^2\theta\right)\sin^{3\gamma-4}\theta.
\end{equation}
In terms of the redshift, we have for a large $\gamma$ and  $\theta\sim90^\circ$:
\begin{equation}
\frac{T_n}{T_n^0}=\left[1+\frac{2}{\gamma}\ln(1+z)\right](1+z)^{-3}\leq e^3\left(1-\frac{1}{\gamma}\right), \quad 
\left(z\geq e^{-1}-1\right),
\end{equation}
where $T_n^0$ is the temperature of normal matter at present, and $e$ is here the base of natural logarithm. 
The temperature of the normal matter reaches its maximum $T^{\rm max}_n\simeq20T_0$ at $z=e^{-1}-1$, or $\theta=90^\circ$, though elsewhere it is almost zero (Fig.\ref{Fig_5}). 
If the temperature of the normal matter and that of CMB are the same at present, then $T^{\rm max}_n\simeq55$K.
In particular, the temperature of the normal matter at the beginning of the universe is precisely zero, 
as expected. 

Consequently, the baryogenesis will continue as long as baryons are in the dark state, 
and it will end when the dark baryons upheave to the cold baryons.  
The genesis of matter in the new cosmology does not date back to the very early universe, but still now going on, since $w_n<0$ yet at present. 
\section{Gravitational constant varying with time\label{VGC}}
From (\ref{NewtonConstant}) and (\ref{EXS}), we have the gravitational constant varying with time: 
\begin{equation}
G=\frac{\sin^2\alpha t}{8\pi\beta^2\eta^2}, \quad \frac{\dot{G}}{G}=\frac{2}{\gamma}H.
\end{equation}
Then, the value of $\gamma$ will be directly obtained by knowing the present values $(\dot{G}/G)_0$ and $H_0$.
The most stringent evaluation of the value $(\dot{G}/G)_0$ is provided by the Lunar Laser Ranging (LLR) \cite{LLR}:
\begin{equation}
\left(\frac{\dot{G}}{G}\right)_0=(4\pm9)\times10^{-13}{\rm y}^{-1}.
\end{equation}
This result suggests $\dot{G}_0>0$: an increasing gravitational constant, 
which favors our new gravity theory, though contrary to the expectations by Dirac and Brans-Dicke.  

With $(\dot{G}/G)_0=4\times10^{-13}{\rm y}^{-1}$ and $H_0=73$kms$^{-1}$Mpc$^{-1}=(13.4{\rm Gy})^{-1}$, we have
\begin{equation}
\gamma=2H_0\left(\frac{\dot{G}}{G}\right)_0^{-1}\simeq370.
\end{equation} 
Accordingly, the variation of the gravitational constant with respect to the redshift $z$ is fairly small: 
\begin{equation}
\frac{G_z}{G_0}=(1+z)^{-2/\gamma}\simeq1-\frac{2}{\gamma}\ln(1+z), \quad G_{10}=0.987G_0, \quad G_{20}=0.984G_0.
\end{equation}
The relation in (\ref{StandardPresentTime}) gives 
\begin{equation}
\frac{dt}{d\theta}=\alpha^{-1}=27.3H_0^{-1}=6.38{\rm Gy}/{\rm degree}, 
\end{equation}
from which the standard age of the present universe for $\theta_0=\arccos\sqrt{2/\gamma}=85.8^\circ$ is 
\begin{equation}
t_0=40.9H_0^{-1}=548{\rm Gy}.
\end{equation}
Furthermore, we find that the phase transition will occur at $t_{\rm x+}=555$Gy, and the universe will stop expanding at $t_{\rm max}=574$Gy. 
Accordingly, the lifetime of the universe will be $1150$Gy. 
The value of the symmetry braking scale $\eta$ and the gauge coupling constant $g$ are obtained from (\ref{NewtonConstant}), (\ref{AB}) and (\ref{DetOfAlpha}):
\begin{equation}
\eta\simeq0.2M_P=2.43\times10^{18}{\rm GeV}, \quad g=\sqrt{\frac{3\pi b}{\gamma}G_0H_0^2}\simeq1.5\times10^{-59},
\end{equation}
where $M_P$ is the Planck mass. The coupling constant $g$ is very small, and we confirm the validity of the assumption $j_n=0$, since due to (\ref{parametrization}), $j_n$ will be proportional to $g$.
\section{Graphics\label{GRP}}
All the graphs in this section assume $\gamma=370$, except for the first one.

\end{document}